\documentclass[a4paper,fleqn,usenatbib]{mnras}

\usepackage[T1]{fontenc}
\usepackage{ae,aecompl}

\usepackage{graphicx}	% Including figure files
\usepackage{amsmath}	% Advanced maths commands
\usepackage{amssymb}	% Extra maths symbols

\title[Generation and decay of the magnetic field]{Impact of continuous particle injection on generation and decay of the magnetic field in collisionless shocks}

\author[M. Garasev, E. Derishev]{
Mikhail Garasev,$^{1, 2}$\thanks{E-mail: garasev@appl.sci-nnov.ru}
Evgeny Derishev$^{1}$
\\
$^{1}$Institute of Applied Physics RAS, 46 Ulyanov Street, 603950 Nizhny Novgorod, Russia\\
$^{2}$Lobachevsky State University of Nizhny Novgorod, 23 Gagarin Avenue, 603950 Nizhny Novgorod, Russia\\
}

\date{Accepted XXX. Received YYY; in original form \today}

\pubyear{2016}

\begin{document}
\label{firstpage}
\pagerange{\pageref{firstpage}--\pageref{lastpage}}
\maketitle

\begin{abstract}
We present numerical simulations of the magnetic field turbulence in collisionless electron-positron plasma with continuous injection of new pairs, which maintains anisotropy in the particle distribution over long time. {With these simulations we follow evolution of a small (and therefore uniform) region in the fluid comoving frame modelling} generation and decay of the magnetic field in shocks, where the upstream is modified by two-photon pair production due to self-absorption of the shock's high-energy radiation. We find that the overall picture of magnetic field build-up is consistent with development of Weibel instability. However, the long-term injection of anisotropic pairs in the upstream leads to formation of large-scale structures in the magnetic field, while the small-scale structures are almost absent. We find that being amplified at the shock front this magnetic field mostly preserves its large spatial scale and then slowly decays in the downstream on a timescale approximately equal to duration of the injection phase. The observed decay of the magnetic field is in exceptionally good agreement with predictions of the so-called phase mixing model. Generation of the long-lived magnetic field in relativistic collisionless shocks with injection-modified upstream explains how they can efficiently produce the synchrotron radiation in Gamma-Ray Bursts.

\end{abstract}


%%%%%%%%%%%%%%%%%%%%%%%%%%%%%%%%%%%%%%%%%%%%%%%%%%

%%%%%%%%%%%%%%%%% BODY OF PAPER %%%%%%%%%%%%%%%%%%%%%%%%

\section{Introduction}

{Relativistic collisionless shocks are the source of observed radiation in Gamma-Ray Bursts (GRBs) and Active Galactic Nuclei (AGNs). Such shocks are thought to be capable of particle acceleration (e.g., 
\citep{AchterbergKirk+,KeshetWaxman}) and of generating strong magnetic field even when they propagate in unmagnetized or weakly magnetized medium (e.g., \citet{MedvedevLoeb}), that naturally makes them efficient sources of synchrotron emission. These predictions have been confirmed by numerical simulations 
(e.g., \citet{GRU1, Spit1, Spit2, DK1}).} 

However, numerical simulations show that either the magnetic field generated at the shock front decays very fast, or -- if the magnetic field is already present in the upstream -- there is no particle acceleration \citep{PICsimulReview}. Thus, {as currently seen in the simulations}, particle acceleration and strong magnetic field never meet each other leaving no room for synchrotron radiation from relativistic shocks. Existing models of emission from GRBs and AGNs require efficient synchrotron radiation in apparent contradiction to what simulations imply. 
A solution to this discrepancy is possibly related to various ways in which the upstream magnetic turbulence could affect the decay of magnetic fields in the downstream (e.g., \citet{Milo1, MedZak}). 

Recently a new model for the structure of relativistic shocks was proposed \citep{Der1}, where the build-up of the magnetic field is a two-stage process. First, the magnetic field is generated in a region ahead of the shock front as a result of massive creation of electron-positron pairs with anisotropic distribution in the comoving frame. This region has a width of the order of the electron cooling length, that is many orders of magnitude larger than the plasma skin depth, so that the anisotropy grows very slowly and is maintained over a very long period on the plasma time scale. Second, the magnetic field is amplified at the shock front and then decays in the downstream. According to the model, the decay length is set by the spatial scale of the turbulent magnetic field formed at the stage of prolonged initial build-up. The goal of the model is therefore to reach the magnetic field decay length comparable to the electron cooling length.

In this Letter we report the results of numerical PIC simulations, which were designed to check theoretical predictions about the magnetic field build-up and decay in the new relativistic shock model. 
Our formulation of the problem of the magnetic field build-up at the shock is different from the standard one. We consider the case where the anisotropy of the particle distribution function is small and is maintained {in the upstream} over a long time through continuous injection of new electron-positron pairs, whereas the standard approach deals with a large anisotropy, which quickly builds up in the existing particle distribution just before the shock front arrives and disappears soon afterwards.  

{In our simulations we follow evolution of a small (and therefore uniform) region in the fluid comoving frame, which starts in the upstream with slow injection of anisotropic pairs, goes through a phase of rapid magnetic field growth due to fast increase of the anisotropy(the shock front), and ends in the downstream with no injection and decaying magnetic field. The shock passage is simulated by hand by changing instantaneously the particle distribution function. }

{We expect (and then confirmed) that generation of the magnetic field is due to Weibel instability. Various aspects of this instability at the linear stage have been analyzed by many authors (e.g., \citet{Mor1, Yal1,LyubEichler}). However, the prolonged quasi-stationary saturation stage of Weibel instability, which is of primary interest in this paper, has so far received no attention.}

In order to reduce the numerical noise and extend the available dynamic range, our simulations were performed in the 2D setup with sub-relativistic particle energies. However, we were not able to explore directly the realistic parameter range for the product of the injection duration and the  plasma frequency. Instead, we use results from simulations in the accessible parameter range to derive the scaling and check if there is any indication that this scaling may break outside of the explored range. Our findings are in agreement with analytic estimates in \citet{Der1}. We do not expect any considerable deviation of our 2D simulations from full 3D simulations, as confirmed by a test 3D run, which agrees well with the 2D simulation for the same parameters. 

The paper organized as follows. In Sect.~\ref{sec:setup} we briefly discuss formulation of the problem and the numerical algorithms, which we used in our simulations. In \ref{sec:injection} we present and discuss the results obtained in our simulations of Weibel instability in the case of long  continuous injection of new particles with anisotropic distribution. In \ref{sec:shock} we describe 
modeling of shock passage through pair plasma with the pre-amplified magnetic field and discuss how the prolonged injection affects build-up and subsequent decay of the magnetic field at the shock. We conclude with discussing astrophysical implications of our results and their relation to other works devoted to generation of the magnetic field in shocks. 

\section{Simulation setup}
\label{sec:setup} 

We perform our numerical simulations of the development of magnetic filaments \citep{W0} using  relativistic PIC-code \textit{EPOCH} \citep{PIC1} in 2D3V geometry. For solving Maxwell's equations this code uses standard finite-difference time-domain method (FDTD) \citep{PIC3}. Equations of motion for the particles are solved with the Boris algorithm. The code was modified to allow injection of new particles (electron-positron pairs) with arbitrary distribution function. The injection scheme randomly distributes particles across the computational domain and ensures that there is no perturbation to momenta, charges, and currents: we simultaneously inject two electron-positron pairs with equal but oppositely directed velocities. We use smooth 3-rd order b-spline shape for computational particles and take 500-5000 particles per cell in different simulations for each species in order to reduce the numerical noise. Low level of noise is necessary both to explore the power spectrum of the magnetic field at the shock and to explore the decay of the magnetic field on long time scales.  

Initial setup of our 2D simulations consists of a box filled with isotropic electron-positron plasma with Maxwellian distribution:
\begin{equation} \label{eq:max}
  f_{e^{-}, e^{+}}(\mathbf{p}) = \frac{N_0}{\sqrt{(2\pi m_\mathrm{e}T)^3}}\exp{\left(-\frac{p^2}{2m_\mathrm{e}T}\right)},
\end{equation}
where $m_\mathrm{e}$ is the electron mass, $T$ the temperature measured in energy units, $p$ the momentum of a particle and $N_0$ the (equal) number densities of electrons and positrons.
The initial temperature was set to 50\,keV.  We used Cartesian 2D grids with uniform mesh ranging in size from $800 \times 800$ to $3200 \times 3200$ cells. Periodic boundary conditions were imposed onto computational domain. The size of a cell in each direction was set to 1 Debye length $r_\mathrm{D} = \sqrt{kT/(8\pi e^2N_0)}$ of an isotropic component. {Here $e$ is the elementary charge.}  

Injection of anisotropic electron-positron component starts at the beginning of each simulation at the moment $t = -t_\mathrm{i}$, continues with a fixed rate, and then stops at $t=0$. The injected plasma has elongated two-temperature distribution function with $T_x = 200$~keV and $T_{y, z} = 50$~keV. The number density of the injected component is
\begin{equation}  \label{eqx}
N_\mathrm{a} (t)= N_\mathrm{0}\delta\begin{cases}
            \frac{\displaystyle t_\mathrm{i} + t }{\displaystyle t_\mathrm{i} },& -t_\mathrm{i} \leq t \leq 0, \\
            1,& \text{ t\, >\, 0,}
     \end{cases}
\end{equation}
where $\delta$ is the total number of injected particles relative to the initial number of particles and $t_\mathrm{i}$ the duration of the injection. We track the evolution for a time up to $15000/\omega_{\mathrm{p}0}$, where $\omega_\mathrm{p}(t) = \sqrt{8\pi e^2(N_\mathrm{0}+N_\mathrm{a}(t))/m_\mathrm{e}}$ is the plasma frequency and $\omega_{\mathrm{p}0} = \omega_\mathrm{p}(-t_\mathrm{i})$ the background plasma frequency. In the paper we use dimensionless time 
\begin{equation}
\tau = \int\limits_{-t_\mathrm{i}}^t\omega_\mathrm{p}\mathrm{d}t-\tau_\mathrm{i},  \qquad 
\tau_\mathrm{i} = \int\limits_{-t_\mathrm{i}}^0\omega_\mathrm{p}\,\mathrm{d}t\,.
\end{equation}
 
{Although our goal is to model relativistic shocks, we used subrelativistic particles. Such a choice guarantees that the magnetic modes in the turbulence prevail over electric ones (as it should be in relativistic shocks). On the other hand, we avoid the well-known problem with relativistic plasmas, where particles produce extra noise through unphysical Cherenkov radiation in vacuum. Even so, the numerical noise was the limiting factor in all our simulations. }

In this work, we analyse two sets of simulations. Simulations from the first set were performed with 500-1500 particles per cell per species for different rates and durations of anisotropic plasma injection. The grid used was $1600 \times 1600$ for the longest injection and $800 \times 800$ in other cases. The  purpose of these simulations was to explore how the injection rate and duration affects the evolution and statistical properties of the generated magnetic field. In the second set of simulations, performed on $1600 \times 1600$ and $3200 \times 3200$ grids, the injection lasted for $500\cdot 2\pi/\omega_{\mathrm{p}0}$ with $\delta=0.5$. After the end of injection we modeled shock passage through the plasma by stretching the particle distribution function (doubling $p_\mathrm{y}$ component of each particle's momentum) and letting the system to evolve. 

\section{Magnetic turbulence in the case of continuous injection}
\label{sec:injection} 

\begin{figure}
	\includegraphics[width=\columnwidth]{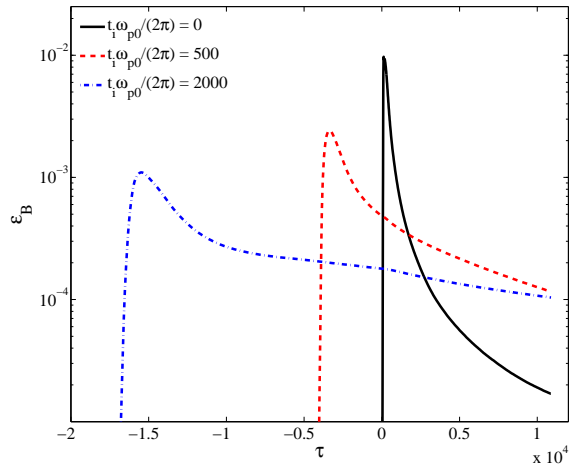}
    \caption{Evolution of magnetization for different durations of injection: instantaneous (solid line), 
$\omega_\mathrm{p0}t_\mathrm{i}/(2\pi) = 500$ (dashed line), and $\omega_{\mathrm{p}0}t_\mathrm{i}/(2\pi)  = 2000$ ({dash}-dotted line). In all cases $\delta = 2$. }
    \label{fig1}
\end{figure}

Examples of the magnetic field evolution observed in our simulations are plotted in Fig.~ \ref{fig1}. The overall picture of the magnetic field build-up agrees with development of Weibel (filamentation) instability. The field starts to grow with relatively small spatial scales and is directed along $z$-axis. The fastest-growing modes initially have their wave-vectors perpendicular to the anisotropy (i.e., aligned with $y$-axis) and the resulting magnetic field islands are strongly elongated in the direction of $x$-axis.

In the case of instantaneous injection our simulations reproduce the result obtained previously by \citep{DK1}. 
The growing magnetic field wipes out anisotropy in the particle distribution, that leads to saturation of the magnetization (the {local} ratio of the energy {density} in the magnetic field to the energy {density} in particles $\varepsilon_B = \langle B^2\rangle/(8\pi E_\mathrm{K}$), where $E_\mathrm{K}$ is a sum of all kinetic energies of particles of both species. Then, the magnetic field starts to decay. The spatial scale of the magnetic field increases at this stage, because the decay rate of long-wave modes is much {smaller} than that of short-wave modes. The maximum level of magnetization in our simulations was about 0.01. This relatively low value is due to sub-relativistic plasma and moderate anisotropy. 

\begin{figure}
	\includegraphics[width=\columnwidth]{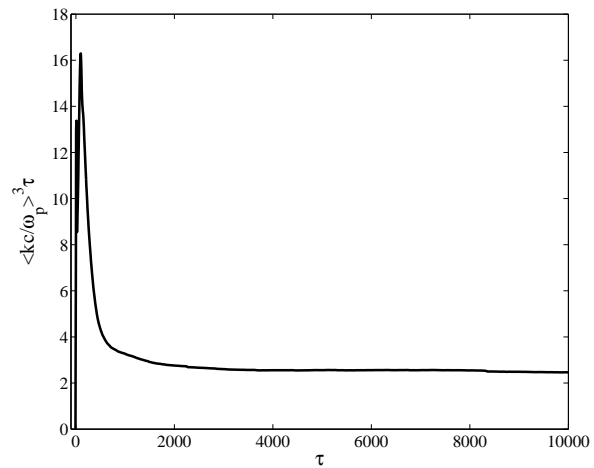}
    \caption{The product $\langle kc/\omega_\mathrm{p} \rangle^3 \tau$ as a function of time for a simulation with instantaneous injection and $\delta = 2$. }
		\label{K3}
\end{figure}

Longer injection results in smaller peak magnetization, but at the same time the magnetic field decays slower. As the time passes, the magnetic field generated by long injection overcomes the magnetic field generated by short injection. Apparently the magnetic field decay follows predictions of the phase mixing model \citep{GRU1,Spit1}. This is illustrated by Fig.~\ref{K3}, which shows that the wavenumber of surviving magnetic perturbations decreases in time in perfect agreement with $\langle k\rangle \propto t^{-1/3}$ law at least up to $t \simeq 10^4 \omega_\mathrm{p}^{-1}$. Numerically, we measure the decrement $\gamma_k = -0.4 (ck)^3/\omega_\mathrm{p}^2$, where the numerical factor cannot be directly compared to theoretical predictions because the cumulative distribution function in our simulations is not thermal. 

\begin{figure}
	\includegraphics[width=\columnwidth]{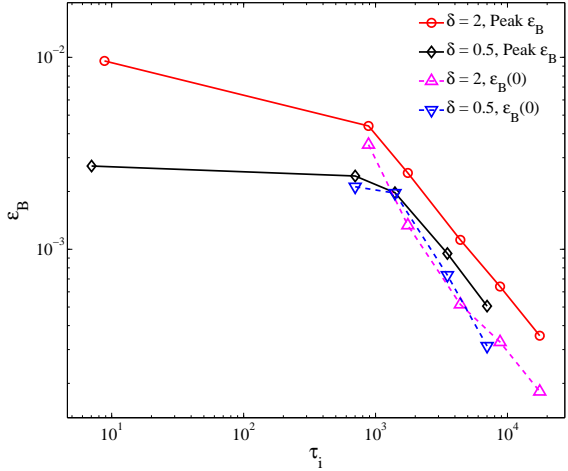}
    \caption{Peak magnetizations $\varepsilon_B$ during the injection phase {and magnetizations at the end of injection phase $\varepsilon_B(0)$ as a functions of the injection duration. } }
		\label{fig2}
\end{figure}

In the case of long injection, where saturation of the magnetic field growth occurs during the injection, we observe a different dynamics. First of all, the longer is the injection, the smaller is the maximum level of magnetization (see. Fig.~\ref{fig2}). On the other hand, the magnetic field decays with a much smaller rate. We attribute this to a larger field spatial scale at the moment of maximum magnetization and to subsequent washing-out of short-wave modes during the rest of injection period.

\begin{figure}
	\includegraphics[width=\columnwidth]{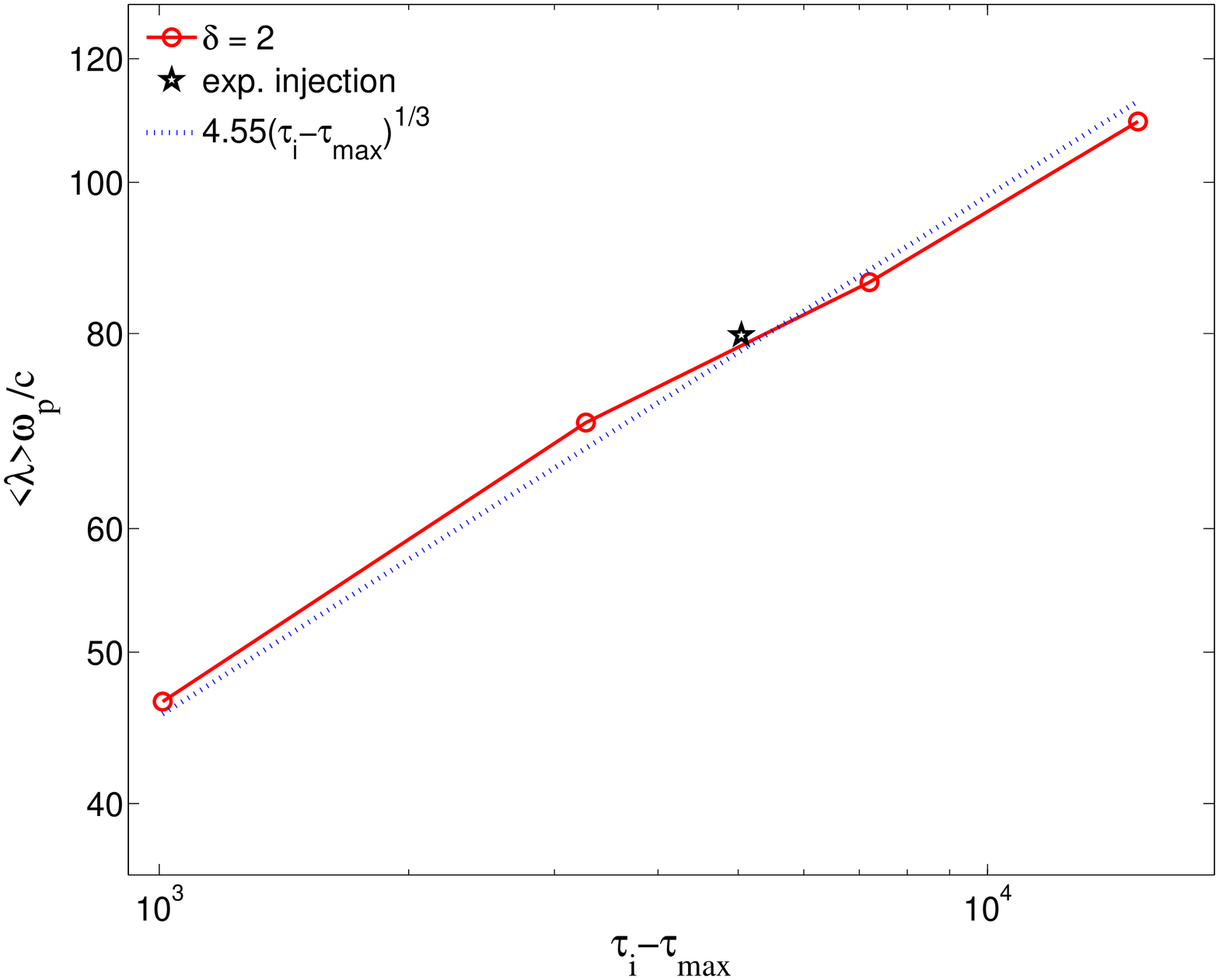}
    \caption{The average wavelength $\langle\lambda\rangle$ at the end of injection as a function of $\tau_\mathrm{i}-\tau_\mathrm{max}$, where $\tau_\mathrm{max}$ corresponds to the time, where maximum of the magnetic field energy is observed. {Star represents a simulation with exponentially growing injection rate, see Eq.~\ref{exp_grow} and the discussion there. }}
    \label{fig3}
\end{figure}

\begin{figure}
	\includegraphics[width=\columnwidth]{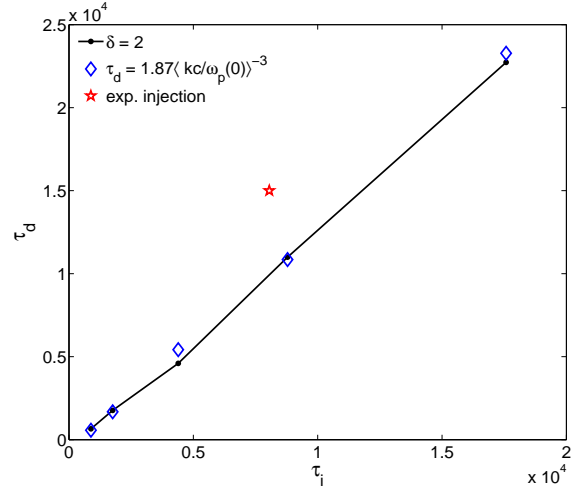}
    \caption{The magnetic field decay timescale as a function of injection duration. Diamonds show the decay timescale predicted theoretically for the average wavenumber of magnetic perturbations measured in these simulations at the end of injection.  {Star represents a simulation with exponentially growing injection rate, see Eq.~\ref{exp_grow} and the discussion there. }}
    \label{fig4}
\end{figure}

Let us define the average spatial scale of the magnetic field as
%\begin{equation}
%<k>_{x, y} = \frac{\int{k_{x, y}|F|^2_{k_x, k_y} \mathrm{d}\,k_y\mathrm{d}\,k_x}}{\int{|F|^2_{k_x, k_y}\mathrm{d}\,k_y\mathrm{d}\,k_x}}
%\end{equation}
\begin{equation}
  \langle\lambda\rangle = \frac{2\pi}{\langle k \rangle}, \qquad 
  \langle k \rangle = \frac{1}{\langle B^2\rangle}\int{kP_k\mathrm{d}\,k}, 
	\end{equation}
	\begin{equation}
\langle B^2\rangle= \int{P_k \mathrm{d}\,k},
\end{equation}
where $k$ are the wavenumbers of magnetic field perturbations and $P_k$ their power spectrum. 
 The average spatial scale at the end of injection for different injection durations is plotted in Fig.~\ref{fig3}.  The dissipation time scale for the magnetic field, defined as  

\begin{equation}
\tau_\mathrm{d} = \frac{\tau_2-\tau_1}{\ln (\varepsilon_\mathrm{B}(\tau_1)/\varepsilon_\mathrm{B}(\tau_2))},
\label{eqt}
\end{equation}
where $\tau_1 = 0.1\, \tau_\mathrm{i}$ and $\tau_2 = 0.25\, \tau_\mathrm{i}$ are two moments of time after injection has stopped, is shown in Fig.~\ref{fig4} and compared to the predictions of the phase mixing model. It is clear that the predictions are in a good agreement with the measured time scales. 

In order to check validity of our results in 3D, we performed a test run with 
$\omega_{\mathrm{p}0} t_\mathrm{i}/(2\pi) = 200$, $\delta=2$, and with 12 particles per cell in $400\times400\times400$ cells computational domain. When this simulation is compared to 2D simulation with the same injection parameters we find exactly the same magnetic field decay time at the end of injection ($\tau_\mathrm{d}\approx 1680$ in 3D vs. $\tau_\mathrm{d}\approx 1760$ in 2D), but the field energy is $\simeq 2.5$ times larger in the 3D simulation, that likely results from slower isotropization of particles in 3D.

{
To test robustness of our results we also performed a simulation with exponentially growing injection rate
\begin{equation}
N_\mathrm{a}(t) = N_\mathrm{0} \delta \begin{cases}
                                  \frac{\mathrm{e}^{t/t_\mathrm{r}}}{t_\mathrm{r}(1-\mathrm{e}^{-t_\mathrm{i}/t_\mathrm{r}})}, & -t_\mathrm{i} \leq t \leq 0, \\
																	1,& \text{ t\, >\, 0,}
																	\end{cases}
\label{exp_grow}
\end{equation}
with $\delta = 2$, $t_\mathrm{i}/(2\pi)  = 1000$, $t_\mathrm{r}  = t_\mathrm{i}/2$, and with 1600 particles per cell in $1600\times1600$ cells computational domain. The average wavelength at the end of the injection phase and the decay time are shown in Figs.~\ref{fig3} and ~\ref{fig4}. It is evident that  constant injection rate represents a conservative scenario. All other things being equal, the exponentially growing injection rate results in a larger magnetization ($\varepsilon_B = 4.0\times10^{-4}$ at the end of injection) and longer decay time.}

\section{Shock modeling}
\label{sec:shock}

The magnetic field, which was generated in plasma at the stage of prolonged pair injection, is later amplified at the shock front. To explore what happens there to the magnetic field, we model rapid build-up of anisotropy at the shock front by instantaneously stretching the particle distribution function for both electrons and positrons along the x-axis, multiplying $p_x$ component of their momenta by a factor of two. The simulation box remains uniform, that implies the actual shock-front width is much larger than the size of the simulation box. {Keeping the simulation box uniform allows us to obtain lower statistical fluctuations in the power spectrum of the magnetic field. Turning on the anisotropy instantaneously instead of more realistic gradual increase creates a bias towards a faster growth of the small-scale modes, thus acting against the effects, which we examine in this paper, and making our results conservative.}

In this set of simulations the initial pre-injection of anisotropic pair plasma continues for $t_\mathrm{i} = 500\times2\pi/\omega_{\mathrm{p}0}$ and the injected particle fraction is $\delta = 0.5$. {To compare the results with the case without prolonged injection we ran a simulation of the shock passage starting with isotropic plasma, whose distribution function is set to repeat the distribution function found at the end of injection phase. }

\begin{figure}
	\includegraphics[width=\columnwidth]{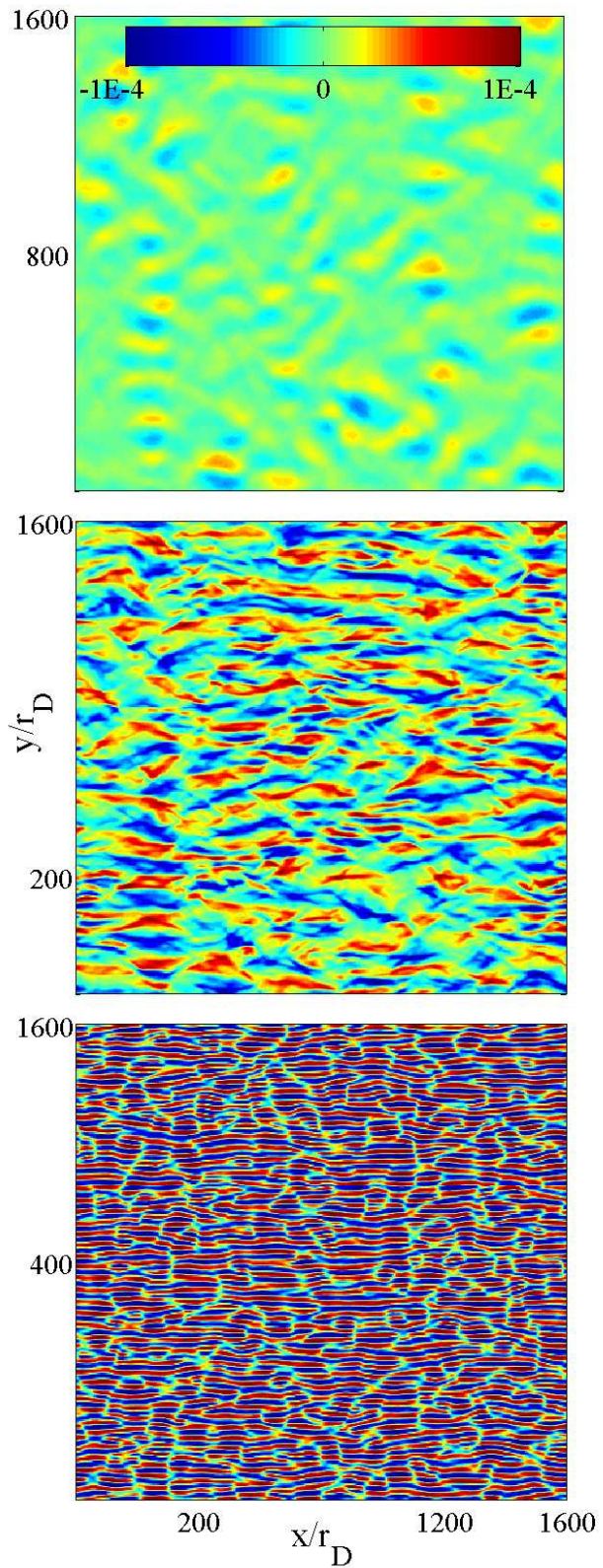}
    \caption{The structure of the magnetic field just before the shock passage at $\tau = 0$ (upper panel) for the simulation with pre-injection ($\omega_{\mathrm{p}0}t_\mathrm{i}/(2\pi)= 500, \delta = 0.5$) and shortly after the passage at $\tau = 100$ (approximately at the maximum of the magnetic field) for the same simulation (middle panel) and for the case without pre-injection (lower panel). }
    \label{fig7}
\end{figure}

In the simulations, we observe that the pre-existing large scale structures of the magnetic field are mostly preserved after the field has been amplified at the shock passage. Small scale structures, which grow by a much larger factor, also become clearly visible, but their contribution is never dominant. And even this contribution is likely to be an artefact of instantaneous anisotropy rise in our simulations. 
The structures of the magnetic field just before the shock and at the moment of maximum magnetization after the shock passage are shown in Fig.~\ref{fig7} and compared to the structure observed at the shock without initial prolonged injection. The power spectra of the magnetic turbulence corresponding to the snapshots from Fig.~\ref{fig7} are presented in Fig.~\ref{fig9}. 

\begin{figure}
	\includegraphics[width=\columnwidth]{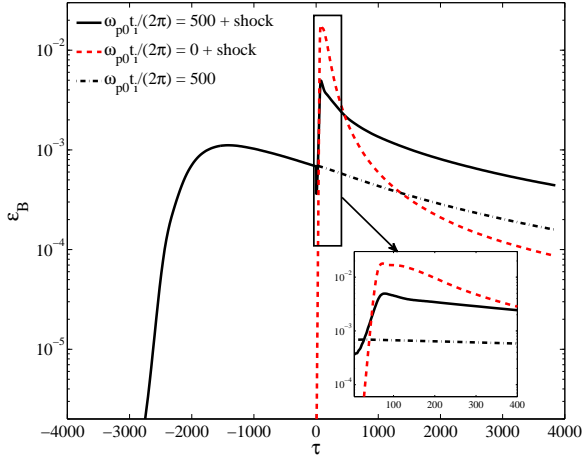}
    \caption{Evolution of the magnetic field at the shock. Solid line: the shock passage is preceded by injection of anisotropic plasma component with $\delta = 0.5$ and $t_\mathrm{i} = 500\cdot2\pi/\omega_{\mathrm{p}0}$. Dashed line: the shock passes through plasma with zero magnetic field (no preceding injection). {Dash-dotted line: the decay of the magnetic field after injection without passage of the shock front.} Inset: rise of the magnetic field at the shock. }
    \label{fig8}
\end{figure}

\begin{figure}
	\includegraphics[width=\columnwidth]{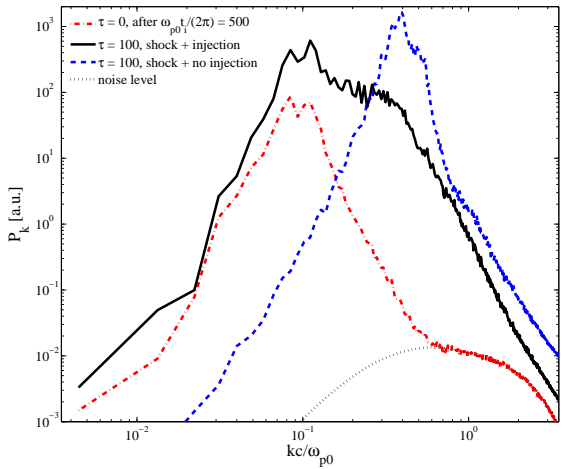}
    \caption{Power spectra $P_k$ of the magnetic field just before the shock passage ($\tau=0$) and shortly after the shock passage at $\tau=100$, that approximately corresponds to the maximum of the magnetic field. }
    \label{fig9}
\end{figure}

Evolution of the magnetization with time during our simulation is shown in Fig.~\ref{fig8}. The simulation includes the injection phase, amplification of the magnetic field at the shock, and the decay phase. This is compared with the same simulation, where -- instead of modeling the shock passage -- we just switched off the injection and let the magnetic field decay on its own.
It is evident from the picture that the magnetic field is quickly amplified at the shock, and that its decay rate after the shock passage rapidly approaches the rate seen in the no-shock case, being somewhat faster at the beginning due to decay of small-scale magnetic field perturbations grown at the shock. 
The magnetic field decay is much slower compared to the case, where Weibel instability starts from zero-field initial conditions, that corresponds to the standard formulation of the problem of magnetic field generation at a shock. The difference is due to dominant contribution of long-lived large-scale modes in the magnetic field turbulence, which are predominantly generated during the injection phase and are further amplified by Weibel instability at the shock. {In contrast, the strength of the seed magnetic field does not directly influence magnetization at the shock front, which saturates due to onset of nonlinear effects as it would be in absence of the seed fluctuations.}

%%%%%%%%%  CONCLUSIONS AND ACKNOWLEDGEMENTS %%%%%%%%%%%%%%%%%%%%
\section{Conclusions}

The results of our simulations demonstrate that Weibel instability -- when it develops in plasma with continuous supply of particles with anisotropic distribution -- leads to generation of large-scale magnetic fields. The typical spatial scale of the generated magnetic field is much larger than $c/\omega_\mathrm{p}$. This scale increases for longer injection as 
$\langle\lambda\rangle \propto (\tau_\mathrm{i}-\tau_\mathrm{max})^{1/3}$ and the magnetic field decay time appears to be approximately equal to the injection duration. 
We show numerically that this scaling holds for injections lasting up to $2\times 10^{4} \omega_\mathrm{p}^{-1}$, and there is no indication that it will break at still longer injections.

We interpret the measured dependence $\langle\lambda\rangle \propto t^{1/3}$ as a proof for the so-called
phase mixing model of the magnetic field decay, which predicts exactly this dependence. Though, the actual numerical coefficient seen in our simulations is somewhat different from the predictions, possibly because the cumulative particle distribution function is not thermal.

The power spectrum of the magnetic turbulence generated during prolonged injection phase declines steeply toward small scales. When we model amplification of the magnetic field at the shock front, we observe that the contribution of large-scale modes to the energy density of the magnetic field remains dominant despite much larger increment for small-scale modes. After amplification at the shock front the large-scale magnetic field survives for a long time in the downstream, explaining efficient synchrotron emission from relativistic shocks. Our results lend support to the shock modification model of \citet{Der1}. 

Although our results are obtained in a simplified model in 2D configurations, we expect similar behaviour in 3D. This was confirmed by a test 3D run, which shows that the decay of the large-scale magnetic field in 3D does not speed up.

Finally, we note an analogy to the results of \citet{Kes1}, where an increase of both magnetic field spatial scale and dissipation length was observed during long-term shock simulations. This corresponds to effective injection timescale becoming progressively larger as the particles are accelerated and penetrate further into the upstream, that leads to generation of larger-scale and more long-lived magnetic field. In view of our results, this analogy suggests that numerical simulations of relativistic collisionless shocks {propagating into unmagnetized medium} will never converge without introducing some dissipation or particle losses.

\section*{Acknowledgements}
The EPOCH code was developed as a part of the UK EPSRC grants EP/G054950/1, EP/G056803/1, EP/G055165/1 and EP/ M022463/1. This work was supported in part by the Government of the Russian Federation Project No. 14.B25.31.0008 (simulation of magnetic turbulence in a weakly anisotropic plasma) and by the Russian Science Foundation grant No 16-12-10528 (simulation of magnetic field amplification at the shock front and its decay in the downstream).

%%%%%%%%%%%%%%%%%%%%%%%%%%%%%%%%%%%%%%%%%%%%%%%%%%



\bibliographystyle{mnras}
\bibliography{weibel_biblio} % if your bibtex file is called example.bib

% Don't change these lines
\bsp	% typesetting comment
\label{lastpage}
\end{document}